\documentclass[a4paper]{article}

\usepackage{setspace}
\usepackage{graphicx}
\usepackage{epstopdf}
\usepackage{color}

\begin{document}

\title{\sffamily Electrically controlled quantum dot based spin current injector}

\maketitle


\author{\sffamily
\Large{\emph{Szabolcs~Csonka}} \footnote{Department of Physics, Budapest University of Technology and Economics, Budafoki u 6-8., 1111 Budapest, Hungary},
\emph{Ireneusz~Weymann}\footnote{
Department of Physics, Adam Mickiewicz University, 61-614 Pozna\'n, Poland\ \ \ \

* Corresponding author\ \ Email: \ \texttt{weymann@amu.edu.pl} \ \
} *, and
\emph{Gergely~Zarand}\footnote{
Dahlem Center, Freie Universit\"at Berlin, Fachbereich Physik,
Arnimallee 14, 14195 Berlin, Germany
}
\footnote{
MTA-BME Quantum Phases Lendulet Research Group, Budapest University of
Technology and Economics, Budafoki ut 8, 1111 Budapest, Hungary}
\vspace{10mm}
}

\begin{abstract}
We present a proposal for a fully electrically controllable quantum
dot based spin current injector.
The device consists of a quantum dot that is strongly coupled to a ferromagnetic
electrode on one side and weakly coupled to a nonmagnetic electrode on the other side.
The presence of ferromagnetic electrode results in an exchange field
that splits the dot level. We show that this exchange-induced
splitting can lead to almost full spin polarization of the current
flowing through the device. Moreover, we also demonstrate that the
sign of the polarization can be changed by the gate or the bias
voltage within a switching time in the nanosecond range. Thus
the proposed device can operate as an electrically controlled, fast
switchable spin current source, which can be realized in various
state-of-the-art nanostructures.
\sffamily

\end{abstract}
\vspace{10mm}


\doublespacing

Spin injection is a central problem in the field of spintronics,
and the improvement of its efficiency and its control
enhances the performance of spin based devices.
Most commonly ferromagnetic electrodes are used as sources of spin polarization,
directly injecting spins from a ferromagnet into a connected device.~\cite{Zutic2004}
However, the spin injection strongly depends on the ferromagnetic-normal interface:
material issues and problems such as conductance mismatch
(\cite{vanSon1987},\cite{Schmidt2000}) make the realization of these
boundaries technologically challenging.
As an interface between the ferromagnet and the normal part  tunnel
barriers are widely used.~\cite{Moodera2010}
The polarization of the injected spin current is then limited by the
polarization of the tunneling electrons,  which -- for a
typical ferromagnet -- is in the range of
30-40\%.~\cite{Meservey1994} Another drawback of this configuration
is that the sign of the current spin polarization can only be changed
by rotating the magnetization  of the ferromagnet, e.g. by applying
an external magnetic field. Therefore, the spin polarization
can be switched only relatively slowly, and in many cases the
application of a local magnetic field is needed.

\begin{figure}[t!]
\centering
 \includegraphics[width=140mm]{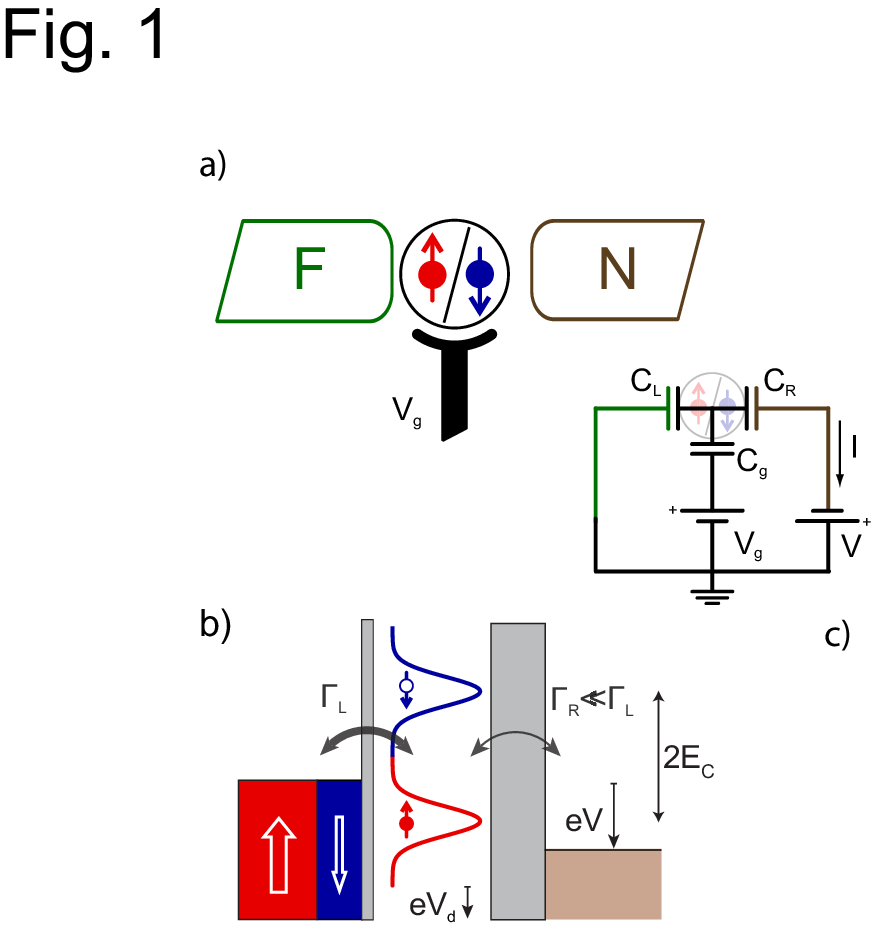}
\caption{
\doublespacing
\sffamily
  a) Device geometry: A quantum dot hosting odd number of
  electrons is placed at the interface of a ferromagnet (F) and normal
  contact (N).  The ground state spin  of the dot (down or up)
  determines the spin polarization of the current. b) Energy diagram
  of the device: The dot is strongly coupled to the ferromagnetic lead
  ($\Gamma_L$) and weakly coupled to the the
  normal lead ($\Gamma_R$). The ferromagnet-induced exchange field
  strongly polarizes the
  two resonance peaks of the dot. For the situation shown, the
  occupied (lower) Hubbard peak is spin-up polarized. The spin
  polarization of the current is generated by the polarization of the
  local density of states in the bias window. $V$ is the applied bias
  voltage and $E_C$ is the charging energy. c) Classical circuit diagram
  of the device: The quantum dot is capacitively coupled to the
  ferromagnetic and normal leads and to the gate electrode with $C_L$,
  $C_R$ and $C_g$, respectively. The gate voltage ($V_g$) allows to
  modify the level position. Due to the capacitances, the applied bias
  voltage ($V$) also shifts the dot level downwards, similar to the gate
  voltage, with a value of $eV_d=-eV C_R/C $, where
  $C=C_L+C_R+C_g$.
}
\label{Fig:1}
\end{figure}

In this paper, we consider a device of different geometry, where the
conventional tunnel barrier is  replaced by a quantum dot (QD)
coupled to a ferromagnetic (F)
and a normal (N) lead (F/QD//N geometry), see {Figure~\ref{Fig:1}a}.
We focus on the case, where the coupling to the ferromagnet
is strong, while the coupling to the normal metal is relatively weak.
Based on accurate numerical renormalization group calculations (NRG)
we  show that then the performance of the spin injection in such a quantum dot
interface is greatly improved:
First, the dot acts as a {\em spin current amplifier}, i.e., the spin
polarization of the injected current highly exceeds that of  the ferromagnetic
electrode, and  may even approach unity.
Second, the polarization of the injected spin current becomes {\em tunable}
by {\em  purely electric means}, either by sweeping the gate voltage or simply by changing the applied bias voltage.
Both ways enable extremely  fast ($\sim$1GHz)
spin polarization switching. In this context, we discuss two
geometries of our spin injection device.
In the first, {\em three-terminal} device,  we control  the spin polarization
by a gate voltage, while in the second, {\em two-terminal} set-up
only the bias voltage is used for control.  While the former geometry
may be more relevant for semiconducting quantum dots,
the latter is more suitable for molecular spintronics applications.

The principle of operation of the proposed spin polarization amplifier is based
on the ferromagnetic exchange field induced level renormalization of
the dot.  When a quantum dot
is coupled to external leads and the number of electrons on the dot is odd,
strong electronic correlations can give rise
to the spin formation and also to the Kondo
effect.~\cite{Goldhaber-Gordon1998,Cronenwett1998}
In case of ferromagnetic leads, quantum fluctuations
renormalize the position of the dot levels
differently for each spin direction
 due to the different spin-up
and spin-down  tunneling rates.
This effective exchange field
then splits the dot levels and suppresses the Kondo
resonance.~\cite{Martinek2003,Choi2004,Pasupathy2004}
An important effect of the exchange field induced splitting
is that the ground state as well as
the local density of states of the dot become highly spin polarized.
This is, in fact, the major ingredient of
the efficient spin polarization amplification discussed
here: due to the exchange splitting of the correlated state,
the quantum  dot will be able to offer spin current polarizations
much higher than the ferromagnet itself.
The second key ingredient is the asymmetry between the left and right contacts:
the coupling to the ferromagnet should be larger than the coupling to
the non-magnetic material to assure that only highly spin polarized states
carry the current in the transport window associated with the bias voltages.
With more symmetrical couplings, one can still operate the device
as an ultra-fast  spin current injector, but polarizations will be reduced.

The proposed F/QD//N structure is quite generic,
and our concept can be realized in various types of
nanostructures using state-of-the-art nanotechnology.
The basic ingredient, i.e. the ferromagnetic contact induced local
exchange field~\cite{Martinek2003,Martinek2005} has already been
demonstrated experimentally in F/QD/F devices, where the QD was
defined using fullerene molecules~\cite{Pasupathy2004}, self-assembled
semiconductor nanocrystals~\cite{Hamaya2007}, or carbon
nanotubes.~\cite{Hauptmann2008,Gaass2011} Very recently, the
ferromagnetic contact induced spin splitting has been observed in
F/QD/N devices, in InAs nanowire based QDs.~\cite{Hofstetter2010}

We note that the problem of current spin polarization in quantum dots coupled to
ferromagnetic and normal leads has already been studied both theoretically
\cite{wang05,weymann06,dalgleish06,souza07,perfetto08,barnas08}
and experimentally~\cite{merchant08,hamaya08}, however, mainly in the weak coupling regime.
In this transport regime the effects related with the ferromagnetic-contact-induced
exchange field are much suppressed and thereby the promising efficiency
and functionality of the spin current injector and spin current
amplifier discussed in this paper are missing.


{\em Theoretical framework.--}
Our model consists of a quantum dot
strongly coupled to the left (L) ferromagnetic lead and weakly coupled
to the right (R) nonmagnetic lead (see Figure~\ref{Fig:1}b).
The strength of the tunnel couplings is described by the tunneling rates
$\Gamma_\alpha$ ($ \alpha=L,R$), with
$\Gamma_L = (\Gamma_{L\uparrow} + \Gamma_{L\downarrow})/2$.
The spin polarization of the ferromagnet is taken into account by
assuming  different  density of states for the spin-up (red) and spin-down
(blue) subbands at the Fermi level, see Figure~\ref{Fig:1}b.  For the
circuit in Figure~\ref{Fig:1}c, the total electrostatic energy of the circuit is
given by,
$E_{\rm tot} = E(n) - e N_R V$. Here the potentials of the left and
right electrodes were explicitly set to $V_L\to 0$ and $V_R\to-V$,
and $N_R$ denotes the numbers of particles on  the right lead.
The first term gives the energy of the quantum dot
\begin{equation}
  E(n) = E_C \left( n- n_g \right)^2
  - \frac{1}{2}\left( C_RV^2 +C_gV_g^2 \right),
\end{equation}
with $n$ the number of particles on the dot,
and
$C_\alpha$ ($V_\alpha$) the capacitances (voltages)
of the left, right, and gate electrodes ($\alpha=L,R,g$),
respectively.
 $E_C = e^2/2C$ denotes the charging energy,
with $C = C_L + C_R + C_g$. The parameter $n_g = (C_gV_g-C_RV)/|e|$
sets  the number of electrons on the dot,
with  $e<0$  denoting the electron charge,  and $V$
the voltage drop on the dot.

At the quantum level, the Hamiltonian of the system can be written as
a sum of three terms, $H = H_L + H_R + T$.
Here the left part $H_L$ describes the dot  and the more strongly coupled
ferromagnetic lead,
\begin{equation}
H_L = H_{0L}
+ E_C(n-n_g)^2 + t_L \sum_{k\sigma} (c_{Lk\sigma}^\dag d_\sigma +
d_\sigma^\dag c_{Lk\sigma}) + {\rm cst.},
\label{eq:F/QD}
\end{equation}
and $H_R = H_{0R} - eN_R V$ describes the right electrode.
In these equations $n$ and $N_\alpha$ are particle number operators,
$n = \sum_\sigma d_\sigma^\dag d_\sigma$, and
$N_\alpha = \sum_{k \sigma} c_{\alpha k\sigma}^\dag c_{\alpha k\sigma}$,
 with $d_\sigma^\dag$ creating a spin-$\sigma$ electron on the dot
and $c_{\alpha k\sigma}^\dag$ being the creation operator of
an electron of wave number $k$ and spin $\sigma$ in electrode $\alpha$.
The Hamiltonian $H_{0\alpha} = \sum_{k\sigma} \varepsilon_{\alpha
k\sigma} c^\dag_{\alpha k\sigma} c_{\alpha k\sigma}$ describes
noninteracting electrons in the leads, while the
last term of $H_L$ describes tunneling between the electrons
in the ferromagnetic lead and the dot with $t_L$ the
corresponding tunneling amplitude.
Finally, the tunneling between the left and right subsystem
is described by
\begin{equation}
T = t_R \sum_{k\sigma} (c_{Rk\sigma}^\dag d_\sigma + d_\sigma^\dag c_{Rk\sigma}),
\end{equation}
with $t_R$ the tunneling amplitude between the right lead and the
dot.

Since the coupling of the dot is assumed to
be much weaker  to the right lead than to  the left lead,
we can perform a perturbative expansion in $T$. In leading order,
we can express  the current in  spin channel $\sigma$ as
\begin{equation}  \label{Eq:Curr}
  I_\sigma(V, V_g) = -\frac{|e|\Gamma_{R}}{\hbar} \int_{-\infty}^\infty d\omega
  A_\sigma(\omega, v_g - v ) \left[ f(\omega) - f(\omega - |e|V)\right]\;,
\end{equation}
where $v_g \equiv C_g V_g/|e|$ and $v = V C_R/|e|$
are the dimensionless gate and bias voltages, respectively (for the sign conventions, see Figure~\ref{Fig:1})
and $A_\sigma(\omega, n_g)$ is the spin dependent
spectral function of the quantum dot coupled  to
the  ferromagnetic lead (F/QD), as described by
Eq.~(\ref{eq:F/QD}). Notice that the second argument of $A$ determines
the position of the dot level. Beside the dimensionless gate voltage,
the level is also shifted by the bias voltage, which has a simple
electrostatic reason (see Figure~\ref{Fig:1}.c). This bias induced shift
is important for the two terminal operation discussed below).
The current  is proportional to
$\Gamma_{R} = 2\pi\rho_R t_R^2$, the tunneling rate
to the right non-magnetic electrode (assumed to have a constant
density of states, $\rho_R$, and $f(\omega)$ denotes the Fermi function.
Equation~(\ref{Eq:Curr})  implies that the
spin polarization of the current,
$P\equiv (I_{\uparrow}-I_{\downarrow})/(I_{\uparrow}+I_{\downarrow})$, is
determined by the difference
of the spectral functions $A_\uparrow$ and $A_\downarrow$
in the energy window between the electro-chemical potential of the left
and right leads. As discussed in the following, the spectral functions
depend strongly on the spin;
due to the ferromagnetic lead induced exchange field,
the two Hubbard peaks get spin polarized with opposite spin
orientation (see Figure~\ref{Fig:1}b).
Large current polarizations can thus be achieved, if the left and
right chemical potentials  are positioned next to one of these peaks.

In order to calculate the spectral function of the F/QD system, we
employ the numerical renormalization group (NRG)
method.~\cite{WilsonRMP75,BullaRMP08,FlexibleDMNRG}
NRG is one of the most powerful tools to study transport through quantum dot
structures, and it captures reliably the ferromagnetic exchange-field induced
spin splitting of the dot levels.~\cite{Martinek2005,Gaass2011} Note, that the accurate non-perturbative NRG approach is essential to describe the problem. Neglecting higher order correlations the spin amplification is lost, the output current polarization can not exceed the F lead polarization.~\cite{Zhu2008}
The NRG calculations were performed for the single impurity Anderson model,
and the ferromagnet was modeled by flat subbands of different density
of states for spin up and spin down electrons.~\cite{Martinek2003,Choi2004}
We assumed parameters typical for quantum dot
structures, and used $E_C / \Gamma_L = 10$, and a spin polarization of the ferromagnet $p=0.4$.
Having determined the spectral functions, we used Eq.~(\ref{Eq:Curr})
to compute  the current.
For the junction capacitances we assumed that $C_L/C_R=2$, and $C_g/C_R=0.1$,
which are typical for semiconducting nanowire QDs.~\cite{Hofstetter2010}

The maximum value of the injected current is given by $I_0=|e|\Gamma_R/\hbar$.
Taking typical semiconducting QD parameters~\cite{Hofstetter2009},
$E_C=5\;$meV, $\Gamma_L=0.5\;$meV, and fulfilling
the assumption of asymmetric couplings, i.e. $\Gamma_R=\Gamma_L/10$,
yields the maximum current in the range of $I_0 \approx 50\;$nA.
However, if $E_C$ and $\Gamma_L$ are larger,
as it is for e.g. carbon nanotube QDs~\cite{Hauptmann2008,Gaass2011},
$I_0$ can be further enhanced. For molecule based QDs~\cite{Pasupathy2004},
where $E_C$ can be especially high ($E_C>100$ meV),
the maximum current can be even in the range of a few $\mu$A's.

\begin{figure}[t!]
\centering
\includegraphics[width=140mm]{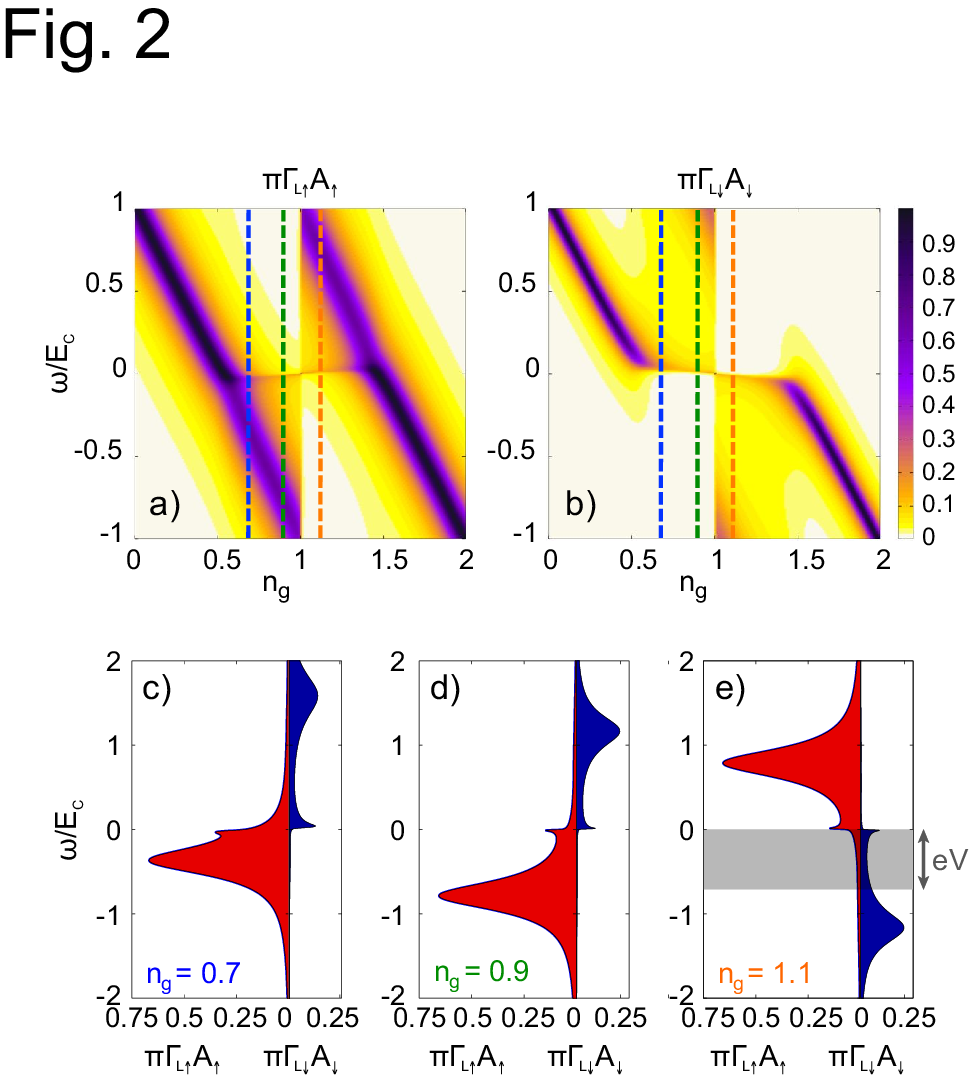}
\caption{
\doublespacing
\sffamily
 Spectral functions of a quantum dot strongly coupled to
  the ferromagnetic lead calculated by NRG:
  Panels a) and b) show the normalized
  zero-temperature spin-up ($A_\uparrow$) and spin-down
  ($A_\downarrow$) spectral functions as a function of energy
  ($\omega$) and effective gate voltage ($n_g$). Panels c), d), and e) show the cross-sections of the
  spectral function at the position of the dashed lines in a) and b)
  i.e. at $n_g=0.7$, $0.9$, $1.1$, respectively. As can be seen, the
  lower (occupied) Hubbard peak is spin-up polarized for $n_g<1$,
  while its polarization changes sign for $n_g>1$. Crossing $n_g=1$
  either by gate or bias voltage induced level shift ($-eV_d$), the
  sign of the current polarization can be reversed. The gray area in
  Panel e) sketches the states contributing to the current at the bias
  voltage $V$. For the NRG calculation we used the
  parameters $p=0.4$, $E_C =0.1D$, $\Gamma_L=0.01D$, and $D\equiv 1$ the half bandwidth.}
\label{Fig:2}
\end{figure}

{\em Spectral functions.--}
The results of our calculations for an unbiased dot ($V=0$)
are shown in \textbf{Figure~\ref{Fig:2}}. Panel a(b)
presents the colorscale plot of the normalized zero-temperature
spin-up (spin-down) spectral function $A_\uparrow(\omega,n_g)$
($A_\downarrow(\omega,n_g)$) with $\omega$  the energy
measured from the Fermi level of the ferromagnet, and $n_g$ is the
effective gate voltage. Upon increasing
$n_g$, the dot's occupation changes from
even to odd and then back to even. The change
in charge occurs at $n_g\approx 1/2$ and  $n_g\approx 3/2$, where
the neighboring charge states with odd and even electrons become
degenerate, and the Hubbard resonances reach the Fermi energy, $\omega=0$.
The region around $n_g=1$ is the Coulomb blockade region with an odd number of electrons
(in our case $\langle n\rangle \approx 1$).
Here additional lines associated with Kondo correlations appear  at  $\omega\approx 0$.
In this regime, the dot level is occupied by a single electron,
and the conduction electrons try to screen
it through the Kondo effect.
This would generate a resonance located normally at the Fermi
level.~\cite{Goldhaber-Gordon1998,Cronenwett1998}
However, because of the ferromagnetic lead, an exchange field acts
on the dot, and splits  and suppresses the Kondo
resonance.~\cite{Martinek2003,Choi2004} This results in smaller
resonances occurring at energies corresponding to the magnitude of the
exchange field. For the spin-up and spin-down orientations the
resonances shift in opposite direction, as clearly seen in
the cross-sections of the colorscale plots, Figs.~\ref{Fig:2}c-e.
The magnitude of the shift is equal to the effective
ferromagnetic exchange field ($B_{\rm exch}$) induced Zeeman splitting,
i.e. $g \mu _B B_{\rm exch}$. Importantly, due to correlation effects,
this exchange field is not constant, but monotonously  changes with $n_g$
(see Figure~\ref{Fig:2}), and  even {\em reverses sign} (see $n_g\approx 1$),
as also demonstrated experimentally.~\cite{Hauptmann2008,Hofstetter2010}
This effect forms the basis of the proposed  electrical spin
polarization control.

Besides the splitting of the Kondo resonance, the ferromagnetic
exchange field has another, even more  dramatic consequence on the spectral
functions: the Hubbard (resonance) peaks become almost fully
spin-polarized. For $n_g<1$ (see Figure~\ref{Fig:2}c,d) the lower/upper
Hubbard peak is spin-up/spin-down polarized and when the exchange
field changes sign (for $n_g>1$) the spin orientation of the Hubbard
peaks is also reversed, see Figure~\ref{Fig:2}e. The proposed spin
current injector is based on this robust polarization of the Hubbard
peaks. If the current flows e.g. in the energy window shown by the
gray stripe in Figure~\ref{Fig:2}e, it is strongly polarized due to the
much larger contribution of the spin-down spectral density.

\begin{figure}[t!]
\centering
\includegraphics[width=140mm]{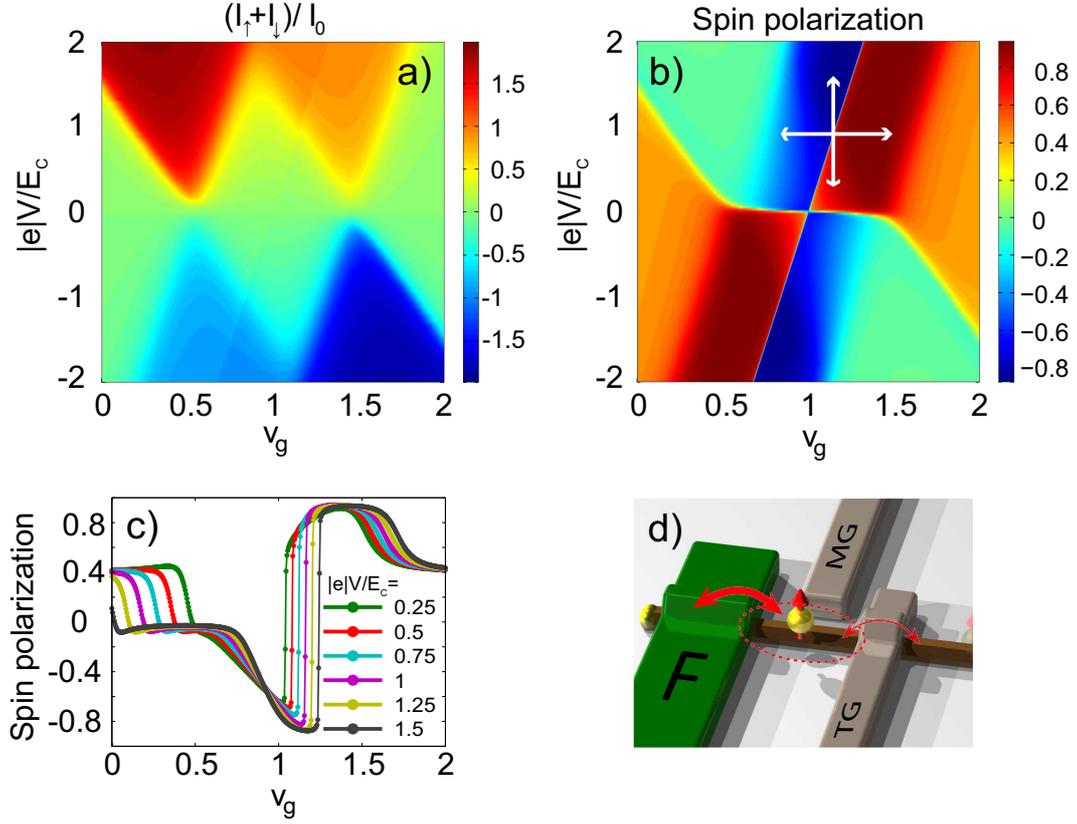}
\caption{
\doublespacing
\sffamily
Three terminal device operation: a) Total current
$I=I_\uparrow+I_\downarrow$ in units of  $I_0 =|e|\Gamma_R/\hbar$
as a function of the dimensionless gate voltage,
$v_g = C_g V_g/|e|$, and applied bias voltage $V$. $I/I_0$ shows the
  expected Coulomb diamond behavior.
 b) The spin polarization of the current for the
  same parameter range. As it is indicated by the white arrows the
  spin polarization can be reversed either by modifying  the level
  position or changing the bias voltage. Panel c) shows the spin
  polarization characteristics as a function of dimensionless gate
  $v_g$ for different bias voltages. Panel d) shows schematic of the
  proposed realization of three terminal device: It consist of a
  nanowire (or a nanotube) strongly coupled to a ferromagnetic lead
  (F) with a top gate (TG) and a middle gate (MG). MG is used to
  change the level position of the dot, while TG defines the tunnel
  barrier with weak coupling. The calculations were performed for
  spectral functions shown in Figure~\ref{Fig:2} with $\Gamma_R/\Gamma_L
  = 0.1$. The capacitances were $C_L/C_R=2$, and $C_g/C_R=0.1$.
} \label{Fig:3}
\end{figure}

{\em Three-terminal operation.--}
The current flowing through the F/QD//N device and its spin
polarization, $P$, are calculated by plugging the NRG spectral functions
into Eq.~(\ref{Eq:Curr}). The results obtained are shown in
Figs.~\ref{Fig:3}a and b. The colorscale plot of the current (Panel a)
shows the expected Coulomb diamond behavior of the quantum
dot. Focusing on the spin polarization (Panel b), the quantum dot acts as a
polarization amplifier: high current polarizations are achieved, which
could strongly exceed the polarization of the ferromagnetic lead
($p=40$\%), and get close even to full polarization. The spin
polarization is strongly enhanced in an extended parameter region,
especially close to the Coulomb peaks (i.e. around $v_g=1/2, 3/2$)
or, for larger bias voltages, outside the Coulomb diamond.
As a spin current source, these regions are also
preferable, since the high polarization is combined with high
amplitude of the current. The sign of the spin polarization is
parallel to the lead polarization in the red region, however, it can
have opposite sign with similarly high amplitude as well (blue region).
This change of polarization is driven by the sign change of the local exchange
field, which induces in turn a polarization change of the occupied Hubbard
peak, see Figure~\ref{Fig:2}d-e.

Changing $v_g$ and thus the occupation
 by the gate voltage along the horizontal arrow
in {Figure~\ref{Fig:3}b}, e.g., the spin polarization of the
current is reversed. This makes the F/QD//N device an efficient
gate-controlled spin current source: the polarization is switchable by
the gate electrode, which can be modulated really fast (over $10\,$GHz
as shown by Nowack {\em et al.}~\cite{Nowack2007})
The speed of the switch can also be limited by
$\Gamma_L$, which is, however typically also in the GHz range
or above, depending on the type of device.

As shown in Figure~\ref{Fig:3}c, the switching of the spin
polarization takes place for a wide range of applied bias voltages with
relatively high up and down polarizations ($P\approx90\%$).
Figure~\ref{Fig:3}d shows a proposed device geometry for the realization
of the gate-controlled spin current source. A ferromagnetic lead couples
strongly to a nanowire (or a carbon nanotube), a middle gate (MG)
tunes the dot level, while a top gate (TG) defines the tunnel barrier
with weak coupling, and the rest of the nanowire serves as a normal
lead. Since the ferromagnetic lead induced local exchange field, its
sign change, fabrication of side and top gates have all been
demonstrated in carbon nanotube~\cite{Hauptmann2008,Gaass2011} or
nanowire~\cite{Hofstetter2010} based structures, the realization of
the proposal is within reach  with state-of-the-art nanofabrication
techniques. Besides the large amplification and the fast gate control, this
geometry could solve the demanding materials issues of spin injection into
carbon nanotubes or nanowires, i.e. the fabrication of proper tunnel
barriers at the interface~\cite{Zutic2004}: coupling a ferromagnetic
lead strongly to these nanoobjects is much simpler, it induces a
ferromagnetic proximity effect that gives rise to highly polarized
spin current, which could be injected through the other weakly coupled
barrier of the quantum dot.

\begin{figure}[t!]
\centering
\includegraphics[width=140mm]{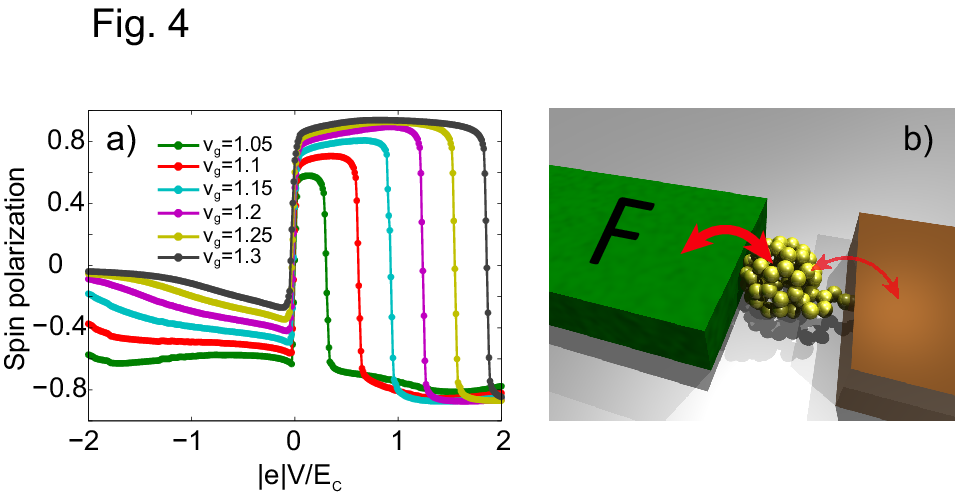}
\caption{
\doublespacing
\sffamily
Two terminal device operation:
a) shows the spin polarization as a function of the bias voltage for
different $v_g$. These are the cross-sections from Figure~3b. The spin
polarization highly exceeds the polarization of the ferromagnetic lead
and it reverses sign for positive bias voltages. Panel b) presents the
proposed realization with asymmetrically coupled molecules.
} \label{Fig:4}
\end{figure}

{\em Two-terminal operation.--}
The F/QD//N system can be also used as a two-terminal spin injector,
where the presence of a gate electrode is \textit{not required}. As it
can be seen in Figure~\ref{Fig:3}b, the value of $v_g$, where the
spin polarization reversal takes place depends on the bias voltage. The
reason for this can be understood based on the simple circuit model
shown in Figure~\ref{Fig:1}c: the applied bias (V) also shifts the dot
level with the value of $eV_d=-eVC_R/C$. As a
consequence, the spin polarization can also be changed by
just varying the bias voltage, e.g. along the vertical arrow shown in
Figure~\ref{Fig:3}b.
We call this two terminal operation, since it does not require the
change of $v_g$
i.e. the change of gate voltage, to reverse the spin polarization.
 As demonstrated in {Figure~\ref{Fig:4}}, the spin polarization as function of the bias voltage reverses in
a wide range of $v_g$. Moreover, polarization reversal
 with highly polarized spin-up and spin-down currents ($|P|>0.6$) can be
generated  even with rather  small bias voltages,  $|V|<0.4 E_C/|e|$
(see the green curve in Figure~\ref{Fig:4}a).

The two terminal operation allows  the realization of the spin injector
using  molecular QD systems,
such as e.g.  modified buckyballs (see inset of
Figure~\ref{Fig:4}b)~\cite{Gruter2005}, where gates are not
necessarily available.
Due to the much larger energy scales of
molecular quantum dots (charging energies of the order of tens - hundreds of meV),
these devices could operate close to room temperature. Furthermore,
molecular quantum dots also allow for the down-scaling of the spin
injector, and support a much faster operation and a larger value of
the injected current, too.
In addition, molecules could also form monolayers, thus the
output current of the spin injector could be enhanced significantly by
the contribution of parallel molecular quantum dots.

Finally, let us comment on the presence of finite temperature, neglected so far.
Since the injected current is almost entirely associated with the Hubbard peaks, the polarization amplification
is effective as long as $T\ll E_C$. However, the polarization switch is
smeared out by a finite temperature, and an effective exchange field
$B_{\rm exch}\gg T$ is required to polarize  the dot spin and reach close to maximum polarizations.
While for quantum dots this may imply below Kelvin device temperatures, for
a molecular device these conditions can be relatively easily satisfied,
and even a room temperature operation may be possible.


{\em Conclusions.--}
In conclusion, we presented a proposal for an efficient spin injector,
which is based on a quantum dot strongly
coupled to a ferromagnetic lead
and weakly coupled to a normal one.
The proposed spin injector has several advantageous
properties: (i) If the ferromagnet-dot coupling is substantially
larger than the normal-dot coupling, then
the output current gets almost fully polarized, since
the dot strongly amplifies the spin polarization of the ferromagnetic
lead. (ii) The polarization of the current can be reversed
purely electrically by a small change of the gate voltage.
The spin switch is thus
 induced \textit{without} modifying the polarization of the
ferromagnetic lead, which  allows an extremely fast polarization
switching.   (iii) The quantum dot based
spin injector can also operate in a two-terminal configuration,
where the current polarization is  reversed by
simply changing  the applied bias voltage.
This allows its implementation in molecular quantum dot
systems as well. (iv) As a molecular electronics device,
the proposed injector
would allow the injection of relatively  large polarized
currents into other nanoscale
objects, and could also be integrated in more complex
nano-devices.

We emphasize again that the basic ingredients of the proposal have been
already demonstrated experimentally in various nanoscale quantum dot systems,
such as carbon nanotubes, semiconductor nanowires/dots or molecular
quantum dots, thus  the realization should be feasible with
state-of-the-art experimental techniques.


\normalsize
We acknowledge support from EU ERC CooPairEnt 258789, FP7 SE2ND
271554, the EU-NKTH GEOMDISS project,
 and Hungarian grants OTKA CNK80991, K73361,
TAMOP 4.2.1./B-09/1/KMR-2010-0002, and EU ERG-239223.
I.W. acknowledges support from the Polish Ministry of Science and
Higher Education through a 'Iuventus Plus' research project for the years 2010-2011,
the Alexander von Humboldt Foundation and the EU grant CIG-303689.
G.Z. acknowledges support from the DFG.


\begin{thebibliography}{10}

\bibitem{Zutic2004}
I. \ifmmode \check{Z}\else \v{Z}\fi{}uti\ifmmode~\acute{c}\else \'{c}\fi{}, J.
  Fabian, and S. Das~Sarma, Rev. Mod. Phys. {\bf 2004,} 76,  323.

\bibitem{vanSon1987}
P.~C. van Son, H. van Kempen, and P. Wyder, Phys. Rev. Lett. {\bf 1978,} 58,  2271.

\bibitem{Schmidt2000}
G. Schmidt, D. Ferrand, L. W. Molenkamp, A. T. Filip and B. J. van Wees, Phys. Rev. B {\bf 2000,} 62,  R4790.

\bibitem{Moodera2010}
G.~M. Jagadeesh S.~Moodera and T.~S. Santos, PysicsToday {\bf 2010,} 63,  46.

\bibitem{Meservey1994}
R. Meservey and P.~M. Tedrow, Physics Reports {\bf 1994,} 238,  173.

\bibitem{Goldhaber-Gordon1998}
D. Goldhaber-Gordon, H. Shtrikman, D. Mahalu, D. Abusch-Magder, U.
Meirav, and M. A. Kastner, Nature (London) {\bf 1998,} 391, 156.

\bibitem{Cronenwett1998}
S.~M. Cronenwett, T.~H. Oosterkamp, and L.~P. Kouwenhoven, Science {\bf 1998}, 281,
  5376.

\bibitem{Martinek2003}
J. Martinek, Y. Utsumi, H. Imamura, J. Barna\'s, S. Maekawa, J. K\"onig, and G. Sch\"on,
Phys. Rev. Lett. {\bf 2003,} 91, 127203.

\bibitem{Choi2004}
M.-S. Choi, D. Sanches, R. Lopez, Phys. Rev. Lett. {\bf 2004} 92, 056601.

\bibitem{Pasupathy2004}
A. N. Pasupathy, R. C. Bialczak, J. Martinek, J. E. Grose, L. A. K. Donev, P. L. McEuen, and D. C. Ralph,
Science {\bf 2004}, 306, 86.

\bibitem{Martinek2005}
J. Martinek, M. Sindel, L. Borda, J. Barna\'s, R. Bulla, J.
K\"onig, G. Sch\"on, S. Maekawa, J. von Delft,
Phys. Rev. B {\bf 2005,} 72, 121302(R).

\bibitem{Hamaya2007}
K. Hamaya, M. Kitabatake, K. Shibata, M. Jung, M. Kawamura, K. Hirakawa,
T. Machida, T. Taniyama, S. Ishida, and Y. Arakawa,
Appl. Phys. Lett. {\bf 2007} 91,  232105.

\bibitem{Hauptmann2008}
J.~R. Hauptmann, J. Paaske, and P.~E. Lindelof, Nature Physics {\bf 2008,} 4,  373.

\bibitem{Gaass2011}
M. Gaass, A. H\"uttel, K. Kang, I. Weymann, J. von Delft, and Ch. Strunk, Phys. Rev. Lett. {\bf 2011,} 107, 176808.

\bibitem{Zhu2008}
Z. G. Zhu, Physics Letters A {\bf 2008,} 372, 695.

\bibitem{Hofstetter2010}
L. Hofstetter, A. Geresdi, M. Aagesen, J. Nygard, C. Sch\"onenberger, and S. Csonka,
Phys. Rev. Lett. {\bf 2010}, 104,  246804.

\bibitem{wang05}
J. Wang and K. S. Chan, D. Y. Xing, Phys. Rev. B {\bf 2005}, 72, 115311.

\bibitem{weymann06}
I. Weymann and J. Barna\'s, Phys. Rev. B {\bf 2006}, 73, 205309.

\bibitem{dalgleish06}
H. Dalgleish and G. Kirczenow, Phys. Rev. B {\bf 2006}, 73, 235436.

\bibitem{souza07}
F. M. Souza, J. C. Egues, and A. P. Jauho, Phys. Rev. B {\bf 2007}, 75, 165303.

\bibitem{perfetto08}
E. Perfetto, G. Stefanucci, and M. Cini, Phys. Rev. B {\bf 2008}, 78, 155301.

\bibitem{barnas08}
J. Barna\'s, I. Weymann, J. Phys. Condens.: Matter {\bf 2008}, 20, 423202.

\bibitem{merchant08}
Christopher A. Merchant and Nina Markovic, Phys. Rev. Lett. {\bf 2008}, 100, 156601.

\bibitem{hamaya08}
K. Hamaya, M. Kitabatake, K. Shibata, M. Jung, S. Ishida, T. Taniyama, K. Hirakawa, Y. Arakawa, and T. Machida,
Phys. Rev. Lett. {\bf 2009}, 102, 236806.

\bibitem{Goldhaber-Gordon2009}
M. Grobis, I. G. Rau, R. M. Potok, H. Shtrikman, and D. Goldhaber-Gordon
Phys. Rev. Lett. {\bf 2008} 100, 246601.

\bibitem{WilsonRMP75}
K. G. Wilson, Rev. Mod. Phys. {\bf 1975,} 47, 773.

\bibitem{BullaRMP08}
R. Bulla, T. A. Costi, and T. Pruschke, Rev. Mod. Phys. {\bf 2008,} 80, 395.

\bibitem{FlexibleDMNRG}
For a description of the code, see:
O. Legeza, C. P. Moca, A. I. T\'{o}th, I. Weymann, G. Zar\'{a}nd,
arXiv:0809.3143 (2008) (unpublished); the code is available at http://www.phy.bme.hu/$\sim$dmnrg/.

\bibitem{Hofstetter2009}
L. Hofstetter, S. Csonka, J. Nygard, and C. Sch\"onenberger, Nature {\bf 2009,} 461, 960.

\bibitem{Nowack2007}
K. C. Nowack, F. H. L. Koppens, Yu. V. Nazarov, and L. M. K. Vandersypen, Science {\bf 2007,} 318, 1430.

\bibitem{Gruter2005}
L. Gr\"uter, F. Cheng, T. T. Heikkil\"a, M. T. Gonzalez, F. Diederich, Ch. Sch\"onenberger and M. Calame,
Nanotechnology {\bf 2005}, 16, 2143.

\end{thebibliography}
\end{document}